\documentclass[twocolumn,english,twocolumn,english,prb,showpacs,superscriptaddress,floats,amsmath,amssymb,floatfix]{revtex4}
\usepackage[T1]{fontenc}
\usepackage[latin1]{inputenc}
\usepackage{textcomp}
\usepackage{amsmath}
\usepackage{graphicx}
\usepackage{amssymb}
\usepackage{esint}
\usepackage{subfigure}

\makeatletter
\@ifundefined{textcolor}{}
{%
 \definecolor{BLACK}{gray}{0}
 \definecolor{WHITE}{gray}{1}
 \definecolor{RED}{rgb}{1,0,0}
 \definecolor{GREEN}{rgb}{0,1,0}
 \definecolor{BLUE}{rgb}{0,0,1}
 \definecolor{CYAN}{cmyk}{1,0,0,0}
 \definecolor{MAGENTA}{cmyk}{0,1,0,0}
 \definecolor{YELLOW}{cmyk}{0,0,1,0}
 }

\@ifundefined{definecolor}
 {\usepackage{color}}{}
\usepackage{epsfig}\usepackage{babel}\allowdisplaybreaks

\makeatother

\usepackage{babel}

\makeatother

\usepackage{babel}

\begin{document}

\title{Thermal drag in spin ladders coupled to phonons}

\author{Christian Bartsch}

\email{c.bartsch@tu-bs.de}

\affiliation{Institut für Theoretische Physik, Technische Universität Braunschweig,
Mendelssohnstrasse 3, D-38106 Braunschweig, Germany}

\author{Wolfram Brenig}

\affiliation{Institut für Theoretische Physik, Technische Universität Braunschweig,
Mendelssohnstrasse 3, D-38106 Braunschweig, Germany}

\date{\today}
\begin{abstract}
We study the spin-phonon drag effect in the magnetothermal transport
of spin-1/2 two-leg ladders coupled to lattice degrees of freedom.
Using a bond operator description for the triplon excitations of the
spin ladder and magnetoelastic coupling to acoustic phonons, we employ
the time convolutionless projection operator method to derive expressions for
the diagonal and off-diagonal thermal conductivities of the coupled
two-component triplon-phonon system. We find that for magnetoelastic
coupling strengths and diagonal scattering rates relevant to copper-oxide
spin-ladders the drag heat conductivity can be of similar magnitude
as the diagonal triplon heat conductivity. Moreover, we show that
the drag and diagonal conductivities display very similar overall
temperature dependences. Finally, the drag conductivity is shown to
be rather susceptible to external magnetic fields.
\end{abstract}


\maketitle

\section{Introduction\label{sec-introduction}}

Understanding spin transport is not only a fundamental issue of quantum
many body physics, but also paramount to future spintronics and quantum
information processing. A new route into \emph{pure }spin transport,
without mobile charge degrees of freedom has been established a decade
by now, with the colossal magnetic heat transport in quasi one-dimensional
(1D) spin ladder materials such as (La,Ca,Sr)$_{14}$Cu$_{24}$O$_{41}$
\cite{Sologubenko2000a,Hess2001a,Kudo2001a} - where the magnetic
contribution to the total thermal conductivity $\kappa$ exceeds the
phonon part substantially - as well as in other 1D spin chain compounds
\cite{Sologubenko2000a,Sologubenko2001a,Hess2007b}. This phenomenon
has led to an upsurge of interest in the non-equilibrium properties
of low-dimensional quantum magnets. Experimentally available data
for the spin transport in ladders is analyzed in terms of Boltzmann
descriptions suggesting very large low-temperature mean-free paths
of several hundred lattice constants \cite{Sologubenko2000a,Hess2001a}.
This remains ill-understood \cite{FHM2007a}.

In this context extrinsic scattering, by impurities and phonons may
play an important role. Phonons by themselves are omnipresent carriers
of heat, which directly interact with spin degrees of freedom, primarily
via magnetoelastic coupling. Usually this interaction is treated as
a source of dissipation of the spin and phonon heat currents \cite{Shimshoni2003a,Rozhkov2005a,Chernyshev2005a,Louis2006a}.
However, another less well studied consequence of a coupled spin-phonon
two-component system exists: namely the off-diagonal effect of the
flow of one of the excitations facilitating the flow of the other
\cite{Gurevich1967a,Boulat2007a,Chernyshev2007a}. This is referred
to as {}``spin-phonon drag,\textquotedblright{} in analogy with electron-phonon
drag discussed for thermoelectric phenomena in metals and semiconductors
\cite{Holstein1964a,Bass1990a,Gurevich1946a,Ziman1946}.

\begin{figure}[tb]
\begin{centering}
\includegraphics[width=0.8\columnwidth]{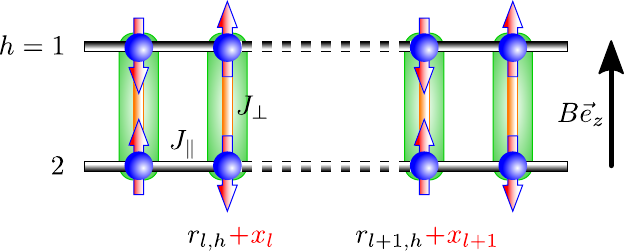} 
\par\end{centering}

\caption{\label{fig1}Spin-phonon coupled ladder model. Red arrows: spin-1/2
moments located at equilibrium positions $r_{l,h}$ on legs $h=1,2$
with longitudinal displacements $x_{l}$ \emph{independent} of $h$.
Dashed leg section: finite longitudinal phonon amplitude $\Delta x_{l}=x_{l}-x_{l+1}\neq0$.
AFM exchange $J_{\parallel(\perp)}$ on black legs (orange rungs).
Green rectangles: rung singlet formation. Spin 'directions': AFM nn-correlations.
$B\vec{e}_{z}$: external magnetic field.}
\end{figure}

Quite recently the theory of spin-phonon drag has been revisited within
a generic two-component model of interacting bosons \cite{Gangadharaiah2010a}.
Within this work, the formal equivalence of two distinct approaches
for the description of drag, namely quasiclassical Boltzmann transport
theory and Kubo linear response formalism, has been laid out, including
several qualitative conclusions. However, a direct application of
the ideas put forward to a more realistic situation is still lacking.
Therefore, in this work we move ahead and study the spin-phonon drag
in a spin-ladder coupled to acoustic phonons. One of our main conclusions
is, that for magnetoelastic coupling strengths typical for cuprate
ladders such as (La,Ca,Sr)$_{14}$Cu$_{24}$O$_{41}$, the drag conductivity
can be of similar magnitude as the magnetic thermal conductivity and
moreover displays a remarkably similar high-temperature dependence.

The outline of this paper is as follows. In section \ref{sec:Spin-Phonon-coupled-ladder}
we describe our model of a two-leg ladder coupled to lattice degrees
of freedom. The spin dynamics of the ladder is treated in the limit
of strong rung coupling, for which we resort to a bond-operator treatment
in subsection \ref{sub:Spin-dynamics-of}. The spin-phonon coupling
is discussed in subsection \ref{sub:Spin-Phonon-Coupling}. In section
\ref{sec:Thermal-Transport-and} we detail our evaluation of the drag rates,
for which we extend the approach of ref. \cite{Gangadharaiah2010a},
by embedding its findings within the time convolutionless projection operator
method to obtain the transport rates. In section \ref{sec:Application-to-Cuprate}
we apply our approach to the specific case of (La,Ca,Sr)$_{14}$Cu$_{24}$O$_{41}$
ladders, contrasting the drag to the diagonal conductivities versus
temperature and spin-phonon coupling strength as well as external
magnetic fields. We conclude in section \ref{sec:Conclusion}.

\section{Spin-Phonon coupled Ladder\label{sec:Spin-Phonon-coupled-ladder}}

For the remainder of this study we focus on the two-leg spin-1/2 ladder
depicted in Fig.~\ref{fig1} with magnetoelastic coupling of the spin
degrees of freedom to lattice vibrations \begin{equation}
H=H_{S}+H_{P}+H_{SP}\,,\label{eq1}\end{equation}
where $H_{S,P,SP}$ refers to the spin, phonon, and spin-phonon parts
of the total Hamiltonian $H$.

\subsection{Lattice Dynamics of the Ladder\label{sub:Lattice-dynamics-of}}

To keep the analysis simple, we refrain from any detailed modeling
of 3D phonon spectra of ladder structures. Rather we assume a spin
ladder with strong coupling across the rungs, both structurally as
well as electronically. In turn intra-rung vibrations are discarded
and we consider only longitudinal inter-rung vibration along the direction
of the legs of the ladder as in Fig.~\ref{fig1}. In turn $H_{P}$
refers to a single \emph{one}-dimensional acoustic phonon \begin{equation}
H_{P}=\sum_{q}\omega_{q}b_{q}^{\dagger}b_{q}^{\phantom{\dagger}}\,,\label{eq2}\end{equation}
where $\omega_{q}{=}\omega_{D}|\text{sin}\left(q/2\right)|$ is the
dispersion of the rung phonon $b_{q,h}^{(\dagger)}$ with a Debye
energy $\omega_{D}$ and $q\in[-\pi,\pi[$. Several comments are in
order regarding this phonon spectrum. First, we stress that other
than for simplicity of later numerical calculations, the formal developments
of our analysis do not depend on the phonon's dimensionality. The
latter primarily affects the low-temperature form of the phononic
heat conduction for $T\ll\omega_{D}$, where it is sensitive to various
scattering mechanism \cite{Klemens1951,Callaway1961}. However, we
will be concerned primarily with the overall behavior of the thermal
conductivity on temperature scales larger than the Debye energy, where
the heat conduction of the pure phonon system is dominated by a constant
specific heat and a scattering rate which increases proportional to
the phonon density, i.e. $\propto T$, independent of dimension.

\subsection{Spin Dynamics of the static Ladder\label{sub:Spin-dynamics-of}}

The spin Hamiltonian of the static ladder reads\begin{equation}
H_{S}{=}\sum_{l}[J_{\perp}\vec{S}_{l,1}{\cdot}\vec{S}_{l,2}+\sum_{h=1,2}(J_{\parallel}\vec{S}_{l,h}{\cdot}\vec{S}_{l+1,h}{-}BS_{l,h}^{z})]\,\label{eq3}\end{equation}
with intra(inter)-rung AFM exchange $J_{\perp(\parallel)}$, external
magnetic field $B$, and periodic boundary conditions (PBC). To describe
the spin excitations of the two-leg ladder for $B=0$ we use the well
established bond-operator representation \cite{jurecka01,sachdev90}
of dimerized spin-$1/2$ systems. Here we briefly recapitulate this
method. The eigenstates of the total spin on a single rung are one
singlet and three triplets. These can be created by the bosonic bond
operators $s_{n}^{\dagger}$ and $t_{\alpha}^{\dagger}$ with $\alpha=x,y,z$
acting on a vacuum $|0\rangle$ by

\begin{eqnarray}
s_{n}^{\dagger}|0\rangle & = & \frac{1}{\sqrt{2}}\left(|\uparrow\downarrow\rangle-|\downarrow\uparrow\rangle\right)_{n}\nonumber \\
t_{x,n}^{\dagger}|0\rangle & = & -\frac{1}{\sqrt{2}}\left(|\uparrow\uparrow\rangle-|\downarrow\downarrow\rangle\right)_{n}\nonumber \\
t_{y,n}^{\dagger}|0\rangle & = & \frac{i}{\sqrt{2}}\left(|\uparrow\uparrow\rangle+|\downarrow\downarrow\rangle\right)_{n}\nonumber \\
t_{z,n}^{\dagger}|0\rangle & = & \frac{1}{\sqrt{2}}\left(|\uparrow\downarrow\rangle+|\downarrow\uparrow\rangle\right)_{n},\label{eq4}\end{eqnarray}
where the first (second) entry in the kets refers to site $1$($2$)
of the rung $n$ of Fig.~\ref{fig1}. On each site we have $[s,s_{\phantom{\alpha}}^{\dagger}]{=}1$,
$[s_{\phantom{\alpha}}^{(\dagger)},t_{\alpha}^{(\dagger)}]{=}0$,
and $[t_{\alpha}^{\phantom{(\dagger)}},t_{\beta}^{\dagger}]{=}\delta_{\alpha\beta}$.
The bosonic Hilbert space has to be restricted to either one singlet
or one triplet per site by the constraint \begin{equation}
s_{n}^{\dagger}s_{n}^{\phantom{\dagger}}+t_{\alpha,n}^{\dagger}t_{n,\alpha}^{\phantom{\dagger}}=1.\label{eq5}\end{equation}
Expressing $H_{S}$ by the bond operators yields a bose gas of singlets
and triplets with two-particle interactions mediated by the inter-rung
coupling $J_{\parallel}$. At $J_{\parallel}{=}0$ these interactions
vanish leaving a sum of purely local rung Hamiltonians, which lead
to a product ground-state of singlets localized on the rungs and a
set of $3^{N}$-fold degenerate triplets. For finite $J_{\parallel}$
the inter-rung interactions can be treated approximately by a linearized
Holstein-Primakoff (LHP) approach \cite{jurecka01,Chubokov89a,chubukov91,Starykh96a,brenig98}.
The LHP method retains spin-rotational invariance and reduces $H_{S}$
to a set of three degenerate massive magnons (triplons) \begin{equation}
H_{S}=\sum_{k}\Omega_{k}a_{\alpha,k}^{\dagger}a_{\alpha,k}^{\phantom{\dagger}}+\mbox{const.}\label{eq6}\end{equation}
with \begin{eqnarray}
t_{\alpha,k}^{\dagger} & = & u_{k}a_{\alpha,k}^{\dagger}+v_{k}a_{\alpha,-k}^{\phantom{\dagger}},\label{eq7}\\
\Omega_{k} & = & J_{\perp}\sqrt{1+2e_{k}}\label{eq8}\\
e_{k} & = & \frac{J_{\parallel}}{J_{\perp}}\cos(k)\label{eq9}\\
u[v]_{k}^{2} & = & \frac{1}{2}\left(\frac{J_{\perp}(1+e_{k})}{\omega_{k}}+[-]1\right).\label{eq10}\end{eqnarray}
The '$[]$'-bracketed sign on the rhs. in (\ref{eq10}) refers to
the quantity $v$ on the lhs.. The spin-gap $\Delta{=}\min\{\Omega_{k}\}$
resides at $k{=}\pi$ with $\Delta{=}\sqrt{J_{\perp}^{2}-2J_{\perp}J_{\parallel}}$.
Note that because of (\ref{eq7}) the ground state $|D\rangle$, which
is defined by $a_{\alpha,k}|D\rangle{=}0$, contains quantum-fluctuations
beyond the pure singlet product-state. To leading order the dispersion
$\Omega_{k}$ is identical to perturbative expansions \cite{Barnes93,Reigrotzki94}.
Beyond the LHP approach triplet interactions and more elaborate consideration
of the constraint (\ref{eq5}) lead to a renormalization of $\Omega_{k}$
and the formation of multi-magnon bound states \cite{jurecka00-2,Kotov98,Kotov98-2,Kotov99,jurecka01-2}.
However, for $J_{\parallel}\ll J_{\perp}$ - which we assume for the
remainder of this work - these renormalizations can be neglected.

To keep $H_{S}$ diagonal at finite magnetic fields $B$ along the
$z$-direction, we have to transform the $x,y,z$-spin projections
of the triplets onto magnetic $z$-quantum numbers $m{=}\pm1,0$ as
usual\begin{equation}
a_{x\{y\},k}^{\dagger}=\frac{1\{i\}}{\sqrt{2}}(a_{1,k}^{\dagger}+\{-\}a_{-1,k}^{\dagger})\,,\hphantom{a}a_{z,k}^{\dagger}=a_{0,k}^{\dagger}\,,\label{eq11}\end{equation}
leading to the final spin Hamiltonian \begin{equation}
H_{S}=\sum_{m,k}\Omega_{m,k}a_{m,k}^{\dagger}a_{m,k}\,,\label{eq12}\end{equation}
with the dispersion \begin{equation}
\Omega_{m,k}=\Omega_{k}+mB\,,\label{eq13}\end{equation}
which splits into three non-degenerate branches at $B\neq0$, according
to the Zeeman energy.

\subsection{Spin-Phonon Coupling\label{sub:Spin-Phonon-Coupling}}

The central ingredient to the spin-phonon drag is some type of coupling
between the triplon and phonon bath. Our work is based on a direct
magnetoelastic coupling along the legs of the ladder. In fact from
Fig.~\ref{fig1} we read off, that the on-leg exchange is subject
to inhomogeneous deviations of the center of mass coordinate of the
dimers from their equilibrium positions by\begin{equation}
J_{||,l,l+1}=J_{\parallel}+\xi\Delta x_{l}\,,\label{eq14}\end{equation}
where $\xi=-\partial J_{||}/\partial a$ with the on-leg lattice constant
$a$, $\Delta x_{l}=(x_{l}-x_{l+1})$, and $\xi$ will usually be
positive. Models for the super-exchange in cuprates \cite{Aronson1991,Cooper1990,Sulewski1990,Kawada1998}
suggest a rather material sensitive range of possible values for $\xi$,
with typically $J_{\parallel}\propto a^{-4}$. From (\ref{eq14})
the spin-phonon part of (\ref{eq1}) follows from its \emph{classical}
version regarding the deviations\begin{equation}
H_{SP}=\sum_{l,h}\xi\Delta x_{l}\vec{S}_{l,h}{\cdot}\vec{S}_{l+1,h}\label{eq:15}\end{equation}
by quantizing the deviations\begin{equation}
x_{q,h}=\sqrt{\frac{\hbar}{2M\omega_{q}}}(b_{-q}^{\dagger}+b_{q})\label{eq16}\end{equation}
in terms of the phonons from subsection \ref{sub:Lattice-dynamics-of}
with the Fourier transform $x_{l}=\sum_{q}e^{iql}x_{q}/\sqrt{N}$.

After some algebra, expressing the spin operators in (\ref{eq:15})
completely analogous to those in (\ref{eq3}) by bond bosons, using
again the LHP approach and dropping all triplon vertices of order
higher than quadratic, we arrive at an expression for $H_{SP}$ in
terms of two-triplon-one-phonon vertices\begin{eqnarray}
\lefteqn{H_{SP}=\sum_{k,q,m,h}(b_{-q,h}^{\dagger}+b_{q,h})\left[\Gamma_{k,q}^{\text{an}}(a_{-m,-k}^{\dagger}a_{m,k+q}^{\dagger}+\right.}\nonumber \\
 &  & \left.\hphantom{aaaaaaa}a_{-m,k}a_{m,-k-q})+\Gamma_{k,q}^{\text{n}}\, a_{m,k+q}^{\dagger}a_{m,k}\right]\hphantom{aaa}\label{eq17}\end{eqnarray}
with \begin{eqnarray}
\Gamma_{k,q}^{\text{an}} & = & -i\,\gamma_{k,q}\, e^{-i(k+(q/2))}\label{eq18}\\
\Gamma_{k,q}^{\text{n}} & = & -2i\,\gamma_{k,q}\,\text{cos}(k+(q/2))\label{eq19}\\
\gamma_{k,q} & = & \frac{g(u_{k}+v_{k})(u_{k+q}+v_{k+q})\text{sin}(\frac{q}{2})}{\sqrt{N|\text{sin}(\frac{q}{2})|}}\label{eq20}\end{eqnarray}
and the coupling constant \begin{equation}
g=\sqrt{\frac{\hbar^{2}}{2M\omega_{D}}}\xi\,.\label{eq21}\end{equation}
We have used $\omega_{q}=\omega_{-q}$. Note that both $\Gamma_{k,q}^{\text{n}}$
and $\Gamma_{k,q}^{\text{an}}$ do not depend on the magnetic quantum
number $m$. The scattering processes in (\ref{eq17}) featuring $\Gamma_{k,q}^{\text{n}}$
correspond to processes where a magnon is scattered under creation
or annihilation of a phonon, the magnon conserves its magnetic quantum
number. These processes will be called \emph{normal} processes in
the following. Terms with $\Gamma_{k,q}^{\text{an}}$ correspond either
to processes, where two magnons of opposite magnetic quantum number
are created (annihilated) along with the annihilation (creation) of
a phonon, or to processes where two magnons of opposite magnetic quantum
number and one phonon are simultaneously created (annihilated). The
former will be called \emph{anomalous} processes and the latter processes
(vacuum fluctuations) are irrelevant for transport.

Next we briefly discuss some very rough orders of magnitude for the
spin-phonon scattering in cuprate ladders. First, in those materials
$J_{\parallel(\perp)}{\gg}\omega_{D}$. Second, both normal and anomalous
processes give rise to final states with at least one triplon, i.e.,
the final DOS is at most $O(1/J_{\parallel}).$ Therefore the parameter
driving perturbation theory is $|g/J_{\parallel}|$. Using (\ref{eq21})
we get $|g/J_{\parallel}|=|(\partial J_{\parallel}/\partial a)(a/J_{\parallel})|/\sqrt{(M/m)(2ma^{2}/\hbar^{2})(\omega_{D})}$
where we have inserted the free electron mass $m$. For the longitudinal
type of dimer phonon, Fig.~\ref{fig1}, $M$ refers to some total
mass of two copper and eight oxygen atoms, i.e., $M/m\sim O(10^{5})$.
For typical cuprate lattice spacings of $a\sim4\text{\AA}$ one has $\hbar^{2}/(2ma^{2})\sim O(100meV)$,
and finally $\omega_{D}\sim O(10meV)$. This yields $|g/J_{\parallel}|=|(\partial J/\partial a)(a/J)|/O(10^{2})$.
With $J\propto a^{-n}$, cf. ref.\cite{Aronson1991,Cooper1990,Sulewski1990,Kawada1998},
it follows that $|g/J_{\parallel}|=O(n/100)$. From $n\sim2\ldots4$,
and given the very approximate nature of the preceding argument,
one may expect $|g/J_{\parallel}|\sim O(10^{-2})\ldots O(10^{-1})$,
which also implies that perturbation theory in terms of $g$ for the
coupled spin-phonon system is well justified.

\section{Thermal Transport and Drag\label{sec:Thermal-Transport-and}}

\subsection{Scattering Rates}

Now we analyze the transport relaxation rates relevant for a Boltzmann
description of the mutual current drag in the coupled spin-phonon
system of the preceding section. Deriving drag rates is not a settled
issue and different routes may be pursued. Direct evaluation of collision
terms have been shown to agree with results from diagrammatic calculations
of the linear-response conductivities \cite{Gangadharaiah2010a}.
Memory functions form another powerful approach \cite{Boulat2007a}.
Here we consider yet an alternative construction of drag rates following
a recent study of electron-phonon coupling in 1D atomic wires \cite{Bartsch2011a}.
Within this construction the time convolutionless (TCL) projection
operator method \cite{TCL1,TCL2,TCL3} is employed to reduce the quantum
dynamics of deviations of occupation numbers in momentum space from
their equilibrium values to a set of rate equations. The later rates
can be interpreted in terms of collision rates of the linearized Boltzmann
equation \cite{Bartsch2010a}. While in ref. \cite{Bartsch2011a}
this approach has been used for the \emph{diagonal} collision rates
of a mixed fermion-boson system, here we will focus on the \emph{off-diagonal}
rates, which describe the drag in a two-component boson-boson system.

For completeness we summarize the main ingredients of the TCL projection
operator method. Given any decomposition of the Hamiltonian $H=H_{0}+V$,
the TCL approach is a perturbative scheme to derive a \emph{closed
system} of equations of motion for a set of 'relevant' quantities
\cite{TCL1,TCL2,TCL3}. In our case the perturbation $V$ refers to the magnon-phonon-coupling
and perturbation theory is applicable in the spirit of the last paragraph
of section \ref{sub:Spin-Phonon-Coupling}, i.e., $g^{2}/J_{\parallel}\ll J_{\parallel},\omega_{D}$.
The main idea is to define a projection operator $\mathcal{P}$ which
maps the system's complete time-dependent density matrix $\rho(t)$
onto a reduced density matrix-like object $\mathcal{P}\rho(t)$ which
only incorporates the dynamics of the relevant variables.

As detailed in \cite{Bartsch2010a}, a proper construction of the
projection operator ${\cal P}$ involves 'coarse graining' of momentum
space. Observable transport processes will be independent of a particular
choice of the graining in the limit of small grain size. Different types
of graining have been discussed in \cite{Bartsch2010a,Bartsch2011a}.
For the present model, one starts off from the deviation of the triplon
(phonon) density at fixed momentum $k$ ($q$) from equilibrium, i.e.
$\Delta_{m,k}^{\text{S}}=a_{m,k}^{\dagger}a_{m,k}-g_{m,k}$ ($\Delta_{q}^{\text{P}}=b_{q}^{\dagger}b_{q}-p_{q}$),
where $g_{m,k}$ ($p_{q}$) are the triplon (phonon) equilibrium Bose
distributions. Using this, the operators\begin{equation}
D_{m,\kappa}^{\text{S}}=\sum_{k\in\kappa}\frac{\rho^{eq}\Delta_{m,k}^{\text{S}}}{\text{Tr}[\rho^{eq}\Delta_{m,k}^{\mathrm{S}\,2}]},\quad D_{\eta}^{\text{P}}=\sum_{q\in\eta}\frac{\rho^{eq}\Delta_{q}^{\text{P}}}{\text{Tr}[\rho^{eq}\Delta_{q}^{\mathrm{P}\,2}]}\label{cb2}\end{equation}
are introduced, where the Greek index $\kappa$ ($\eta$) labels a
grain in the magnon's (phonon's) momentum space and $\rho^{\mathrm{eq}}$
corresponds to the equilibrium density matrix, i.e., $\text{Tr}[\rho^{\mathrm{eq}}a_{m,k}^{\dagger}a_{m,k}]=g_{m,k}$
and $\text{Tr}[\rho^{\mathrm{eq}}b_{q}^{\dagger}b_{q}]=p_{q}$. From
eqn. (\ref{cb2}), the projected density matrix is defined \begin{equation}
\mathcal{P}\rho(t)=\rho^{\mathrm{eq}}+\sum_{m,\kappa}\frac{D_{m,\kappa}^{\text{S}}d_{m,\kappa}^{\text{S}}(t)}{N_{m,\kappa}}+\sum_{\eta}\frac{D_{\eta}^{\text{P}}d_{\eta}^{\text{P}}(t)}{N_{\eta}}\,,\label{cb3}\end{equation}
where $N_{m,\kappa}$ ($N_{\eta}$) are the number of states per grain
and $d_{m,\kappa}^{\text{S}}(t)$ ($d_{\eta}^{\text{P}}(t)$) are
the on-grain density deviations \begin{equation}
\begin{array}{lcl}
d_{m,\kappa}^{\text{S}}(t)=\text{Tr}[\Delta_{m,\kappa}^{\text{S}}\rho(t)] &  & d_{\eta}^{\text{P}}(t)=\text{Tr}[\Delta_{\eta}^{\text{P}}\rho(t)]\\
\\\Delta_{m,\kappa}^{\text{S}}=\sum_{k\in\kappa}\Delta_{m,k}^{\text{S}} &  & \Delta_{\eta}^{\text{P}}=\sum_{q\in\eta}\Delta_{q}^{\text{P}}.\end{array}\label{cb1}\end{equation}
These density deviations form the set of relevant variables.

To leading order in $V$, i.e., $g$, and introducing the vector $\mathbf{d}(t)=(\mathbf{d}^{\textrm{S}}(t),\mathbf{d}^{\textrm{P}}(t))=(d_{m,\kappa}^{\text{S}}(t),d_{\eta}^{\text{P}}(t))$
of the relevant variables, the TCL projection approach now results
in a system of rate equations which couples all magnon and phonon
occupation number deviations as\begin{equation}
\dot{\mathbf{d}}(t)=\mathbf{R}(t)\mathbf{d}(t),\quad\mathbf{R}(t)=\left[\begin{array}{ll}
\mathbf{R}^{\text{S}}(t) & \mathbf{R}^{\text{P-S}}(t)\\
\mathbf{R}^{\text{S-P}}(t) & \mathbf{R}^{\text{P}}(t)\end{array}\right]\,.\label{cb4}\end{equation}
The $2\times2$ block-structured rate matrix allows for a direct interpretation
in terms of momentum redistribution. While the diagonal blocks $\mathbf{R}^{\text{S(P)}}$
account for transitions between momentum grains \emph{within} the
triplon (phonon) sector of the Hilbert space due to the presence of
phonons (triplons), the off-diagonal blocks $\mathbf{R}^{\text{P-S(S-P)}}$
quantify a momentum pickup of the phonons (triplons) from the triplons
(phonons), i.e., mutual \emph{drag.}

Analytic expressions for $\mathbf{R}^{\mathrm{S-P(P-S)}}$ to leading
order in $g$ are evaluated in close analogy to \cite{Bartsch2010a,Bartsch2011a}.
For the drag rates, and after some algebra, we get\begin{eqnarray}
\lefteqn{R_{\eta,(m,\kappa)}^{\text{S-P}}(t)=\sum_{i\in\kappa,j\in\eta}\frac{2}{\hbar^{2}N_{m,\kappa}}\int_{0}^{t}\text{d}\tau}\label{dragratemph}\\
 &  & \left[\vert\Gamma_{i,-j}^{\text{n}}\vert^{2}(1{+}g_{m,i-j}{+}p_{m,j})\text{cos}(\frac{\varepsilon_{m,i}{-}\varepsilon_{m,i-j}{-}\omega_{j}}{\hbar}\tau)\right.\nonumber \\
 &  & +\vert\Gamma_{i,j}^{\text{n}}\vert^{2}(g_{m,i+j}{-}p_{j})\text{cos}(\frac{\varepsilon_{m,i}{-}\varepsilon_{m,i+j}{+}\omega_{j}}{\hbar}\tau)\nonumber \\
 &  & +4\vert\Gamma_{i,-j}^{\text{an}}\vert^{2}(g_{-m,j-i}{-}p_{j})\text{cos}(\frac{\varepsilon_{m,i}{+}\varepsilon_{-m,j-i}{-}\omega_{j}}{\hbar}\tau)\nonumber \\
 &  & \left.+4\vert\Gamma_{i,j}^{\text{an}}\vert^{2}(1{+}g_{-m,-i-j}{+}p_{j})\text{cos}(\frac{\varepsilon_{m,i}{+}\varepsilon_{-m,-i-j}{+}\omega_{j}}{\hbar}\tau\right]\nonumber \end{eqnarray} 
and\begin{eqnarray}
\lefteqn{R_{(m,\eta),\kappa}^{\text{P-S}}(t)=\sum_{i\in\kappa,j\in\eta}\frac{2}{\hbar^{2}N_{\kappa}}\int_{0}^{t}\text{d}\tau}\label{dragratephm}\\
 &  & \left[\vert\Gamma_{j,-i}^{\text{n}}\vert^{2}(g_{m,j-i}{-}g_{m,j})\text{cos}(\frac{\varepsilon_{m,j}{-}\varepsilon_{m,j-i}{-}\omega_{i}}{\hbar}\tau)\right.\nonumber \\
 &  & +\vert\Gamma_{j,i}^{\text{n}}\vert^{2}(g_{m,i+j}{-}g_{m,j})\text{cos}(\frac{\varepsilon_{m,j}{-}\varepsilon_{m,j+i}{+}\omega_{i}}{\hbar}\tau)\nonumber \\
 &  & +4\vert\Gamma_{j,-i}^{\text{an}}\vert^{2}(1{+}g_{-m,i-j}{+}g_{m,j})\text{cos}(\frac{\varepsilon_{m,j}{+}\varepsilon_{-m,i-j}{-}\omega_{i}}{\hbar}\tau)\nonumber \\
 &  & \left.+4\vert\Gamma_{j,i}^{\text{an}}\vert^{2}(1{+}g_{-m,-i-j}{+}g_{m,j})\text{cos}(\frac{\varepsilon_{m,j}{+}\varepsilon_{-m,-i-j}{+}\omega_{i}}{\hbar}\tau)\right].\nonumber \end{eqnarray}
Note that these expressions are valid at \emph{all} times. Only in
the long-time limit they will display the familiar 'energy-conserving'
$\delta$-functions. In each of (\ref{dragratemph}) and (\ref{dragratephm}),
the first two rate terms result from normal processes and the remaining
two from anomalous processes, where the last one corresponds to a
simultaneous creation (annihilation) of two magnons and one phonon
and is forbidden in the long-time limit because of energy conservation.

For the diagonal rates $\mathbf{R}^{\mathrm{S(P)}}$, expressions
similar to (\ref{dragratemph},\ref{dragratephm}) can be derived
in principle. However, for the remainder of this paper we will follow
a different route. Namely, in addition to the triplon-phonon scattering,
we will include all other potentially relevant diagonal relaxation
mechanisms not contained in our model on a phenomenological basis.
This includes impurity scattering, triplon-triplon interactions beyond
eqn. (\ref{eq6}), and phonon anharmonicities. We assume these rates
to be diagonal $R_{(m,\kappa),(n,\eta)}^{\text{S}}{\approx}-\delta_{\kappa\eta}\delta_{mn}/\tau_{m,\kappa}^{\text{S}}$
and $R_{k,i}^{\text{P}}{\approx}-\delta_{\kappa\eta}/\tau_{\kappa}^{\text{P}}$,
with relaxation times $\tau_{m,\kappa}^{\text{S}}$ and $\tau_{\eta}^{\text{P}}$
which may still depend on temperature, the momentum grain index, as
well as the triplon's magnetic quantum number.

\subsection{Thermal Conductivity}

Following the standard Chapman-Enskog procedure \cite{Jaeckle1978}
we now set up the linearized Boltzmann equation. To that end we introduce
additional 'vectors'\begin{align}
\mathbf{n}=(g_{m,\kappa},p_{\eta})\hphantom{aaaaa}\textrm{equilibrium distributions}\hphantom{\,,}\nonumber \\
\boldsymbol{\varepsilon}=(\varepsilon_{m,\kappa},\omega_{\eta})\hphantom{aaaaaaaaaaaaaaaaaaa}\textrm{energies}\hphantom{\,,}\nonumber \\
\mathbf{v}=(\partial_{k}\varepsilon_{m,\kappa},\partial_{k}\omega_{\eta})=(v_{m,\kappa},u_{\eta})\hphantom{aa.a}\textrm{velocities}\,,\label{n1}\end{align}
using the triplon and phonon subsystem indices $m,\kappa$ and $\eta$.
To simplify notations, the triplon indices will also be subsumed
into a single Greek letter $\xi$. For slow spatial and time variations,
i.e., in the hydrodynamic regime, and to leading order, the Chapman-Enskog
expansion links the change of the distribution function to its deviation
from equilibrium via the rates of eqn. (\ref{cb4}) by\begin{equation}
-\frac{\partial n_{\alpha}}{\partial\varepsilon}\frac{\varepsilon_{\alpha}}{T}\left(\mathrm{v}_{\alpha}\partial_{x}T\right)=\sum_{\beta}R_{\alpha\beta}d_{\beta}\,.\label{cb5}\end{equation}
Since the heat current density is $\vec{j}=\frac{1}{Ga}\sum_{\alpha}\varepsilon_{\alpha}\vec{\mathrm{v}}_{\alpha}d_{\alpha}$,
where $G$ corresponds to the number of grains and $a$ is the lattice
constant, the thermal conductivity follows from (\ref{cb5}) as\begin{equation}
\kappa=\frac{1}{GaT}\sum_{\alpha,\beta}\mathrm{v}_{\alpha}\varepsilon_{\alpha}R_{\alpha\beta}^{-1}\varepsilon_{\beta}\mathrm{v}_{\beta}\frac{\partial n_{\beta}}{\partial\varepsilon}\,\label{thermcond}\end{equation}
with the $2\times2$-block inverse collision term matrix $R_{\alpha\beta}^{-1}$.
In turn the thermal conductivity decomposes into four parts\begin{equation}
\kappa=\kappa_{\text{S}}+\kappa_{\text{P}}+\kappa_{\text{P-S}}+\kappa_{\text{S-P}}\,,\label{totcond}\end{equation}
which can be identified as the conventional magnon and phonon conductivities
$\kappa_{\text{S}}$, $\kappa_{\text{P}}$, which are \emph{diagonal},
i.e., $\kappa=\eta$,\emph{ }and two drag conductivities $\kappa_{\text{P-S}}$,
$\kappa_{\text{S-P}}$, which are \emph{off-diagonal}, i.e., $\kappa\neq\eta$.
Assuming that $R^{\text{S}}$, $R^{\text{P}}$ $\gg$ $R^{\text{S-P}}$,
$R^{\text{P-S}}$, the inverse collision term simplifies, and we obtain
the usual diagonal conductivities\begin{eqnarray}
\kappa_{\text{S}} & = & -\frac{1}{Ga}\frac{1}{T}\sum_{\xi}v_{\xi}^{2}\varepsilon_{\xi}^{2}\tau_{\xi}^{\text{S}}\frac{\partial g_{\xi}}{\partial\varepsilon}\nonumber \\
\kappa_{\text{P}} & = & -\frac{1}{Ga}\frac{1}{T}\sum_{\eta}u_{\eta}^{2}\omega_{\eta}^{2}\tau_{\eta}^{\text{P}}\frac{\partial p_{\eta}}{\partial\omega}\,,\label{diagcond}\end{eqnarray}
where $\xi$ ($\eta$) runs over the triplon (phonon) grains only.
The drag conductivities read\begin{eqnarray}
\kappa_{\text{S-P}} & = & -\frac{1}{Ga}\frac{1}{T}\sum_{\eta,\xi}(v_{\xi}\varepsilon_{\xi}\tau_{\xi}^{\text{S}})(u_{\eta}\omega_{\eta}\tau_{\eta}^{\text{P}})\frac{\partial g_{\xi}}{\partial\varepsilon}R_{\eta,\xi}^{\text{S-P}}\nonumber \\
\kappa_{\text{P-S}} & = & -\frac{1}{Ga}\frac{1}{T}\sum_{\xi,\kappa}(v_{\xi}\varepsilon_{\xi}\tau_{\xi}^{\text{S}})(u_{\kappa}\omega_{\kappa}\tau_{\kappa}^{\text{P}})\frac{\partial p_{\kappa}}{\partial\omega}R_{\xi,\kappa}^{\text{P-S}}.\,\,\,\,\,\,\,\,\,\,\,\,\,\label{dragcond}\end{eqnarray}
Using eqns. (\ref{dragratemph},\ref{dragratephm}), and after some
algebra one can show that $\kappa_{\text{S-P}}=\kappa_{\text{P-S}}\equiv\kappa_{\text{D}}$/2.
As a consistency check, we note that eqns. (\ref{dragcond}) are identical
to those derived in ref. \cite{Gangadharaiah2010a}, using however
different methods.

\section{Application to Cuprate Ladders\label{sec:Application-to-Cuprate}}

In this section we will perform a numerical evaluation of (\ref{dragratemph},\ref{dragratephm})
and (\ref{diagcond},\ref{dragcond}), to provide a semi quantitative
estimate of the size of the drag as compared to the diagonal conductivities
using a set of parameters which might be of relevance to the cuprate
spin ladder La$_{5}$Ca$_{9}$Cu$_{24}$O$_{41}$ (LCCO) \cite{Hess2001a,Hess2005a,Hess2006a,Otter2012a}.

For the remainder of this work we will assume the diagonal scattering
times to be temperature dependent but momentum independent. First
we consider \emph{reduced} quantities $G_{\text{S,P,D}}$ for the conductivities
of the triplons, phonons, and drag. These are obtained by stripping
$\kappa_{\text{S,P,D}}$ from their dependence on the diagonal scattering
times $\tau^{\text{S(P)}}$ and the scale of the spin-phonon drag time
$\tau^{\text{D}}$ \begin{equation}
G_{\text{S(P)}}=\frac{\kappa_{\text{S(P)}}}{\tau^{\text{S(P)}}}\,,\hphantom{aaa}G_{\text{D}}=\frac{\tau^{\text{D}}\kappa_{\text{D}}}{\tau^{\text{S}}\tau^{\text{P}}}\,,\hphantom{aaa}\tau^{\text{D}}\equiv\frac{\hbar J_{\parallel}}{g^{2}}\,.\label{g}\end{equation}
These reduced quantities allow to compare the relative size of the
various conductivities versus temperature $T$ and magnetic field
$B$, independent of their mean-free paths and involve the parameters
$J_{\parallel}$ and $\omega_{D}$ only. Using $J_{\perp}\approx \text{k}_{\text{b}}\,1200\,\text{K}$
\cite{Hess2001a,Hess2005a,Hess2006a,Otter2012a}, the former is set
to $J_{\parallel}=0.45J_{\perp}$, which approximately captures the
spin gap $\Delta_{T}$ of the LCCO ladders which is $\Delta_{T}\approx \text{k}_{\text{b}}\,380\,\text{K}$.
For the phonons we chose a Debye energy of $\omega_{\text{D}}\approx0.32J_{\perp}$.
For these parameters only the normal processes in eqn. (\ref{eq17})
contribute to the scattering rates in eqns. (\ref{dragratemph},\ref{dragratephm}).
Finally, only the long-time limit of the latter equations will be
used in the following discussion.

\begin{figure}[tb]
\centering{} \includegraphics[width=0.8\columnwidth]{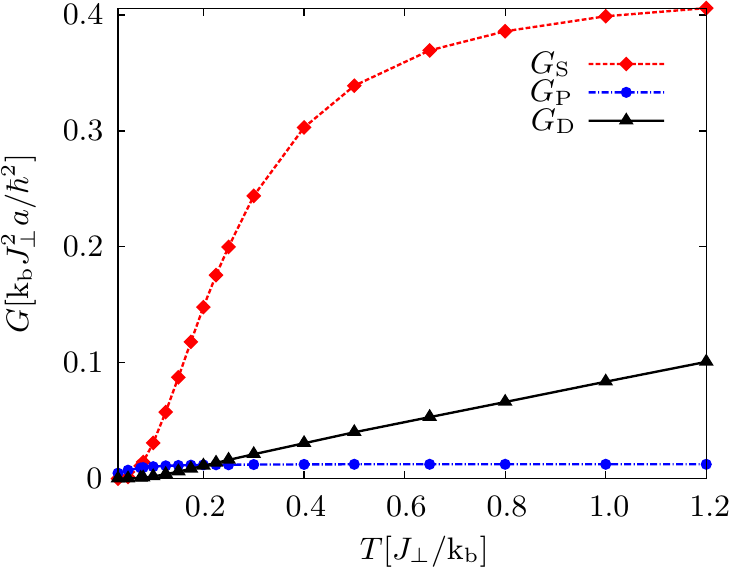} \caption{\label{fig2}Reduced thermal conductivities $G_{\text{S}}$, $G_{\text{P}}$
and $G_{\text{D}}$ from eqn. (\ref{g}) versus temperature at $B=0$.
The drag contribution $G_{\text{D}}$ is of the same order of magnitude
as $G_{\text{S}}$. While $G_{\text{S}}$ and $G_{\text{P}}$ become
constant for large $T$, $G_{\text{D}}$ grows linearly in this limit.
Parameters: $J_{\parallel}=0.45J_{\perp}$, i.e., gap $\approx0.3J_{\perp}$
for $B=0$, $\omega_{D}\approx0.3\, J_{\perp},N=400$.}
\end{figure}

Fig.~\ref{fig2} displays $G_{\text{S,P,D}}$ versus temperature at $B=0$.
For the numerical evaluation we have checked, that the results are
independent of the coarse graining procedure in the limit of small
grains and we have chosen system sizes, typically $N=400$ rungs,
such that finite size effects remain negligible on the scale of plots
discussed in the following. Fig.~\ref{fig2} shows that the ratio
between the reduced triplon and drag conductivities is of $O(10)$
for typical temperatures within the range depicted. With eqn. (\ref{g}),
this implies, that for the triplon and drag conductivities to be of
comparable size, $g^{2}\tau^{\text{P}}/(\hbar J_{\parallel})\sim O(10)$
should hold. As we will show later, this is not unrealistic for LCCO.
The temperature dependence of $G_{\text{S}}$ and $G_{\text{P}}$ is as expected,
i.e., at $T\ll J$, $G_{\text{S}}$ shows exponential Arrhenius behavior,
while $G_{\text{P}}$ increases linearly for $T\ll\omega_{D}$ due to the
$1$D phonon dispersion. The latter is only visible for the very low
temperature scales of Fig.~\ref{fig2}. For $T\gg J$ and $\omega_{D}$,
both quantities $G_{\text{S}}$ and $G_{\text{P}}$ saturate. $G_{\text{D}}$, on the other
hand, \emph{increases linearly} with $T$ in the intermediate and
high temperature range depicted. This is an effect, arising from the
combined high-temperature behavior of the sums of Bose functions in
both drag rates (\ref{dragratemph},\ref{dragratephm}), and the derivatives
$(\partial g/\partial\varepsilon)$ and $(\partial p/\partial\omega)$
in (\ref{dragcond}).

\begin{figure}[tb]
\begin{centering}
\subfigure[]{\label{fig3a}
\includegraphics[width=0.8\columnwidth]{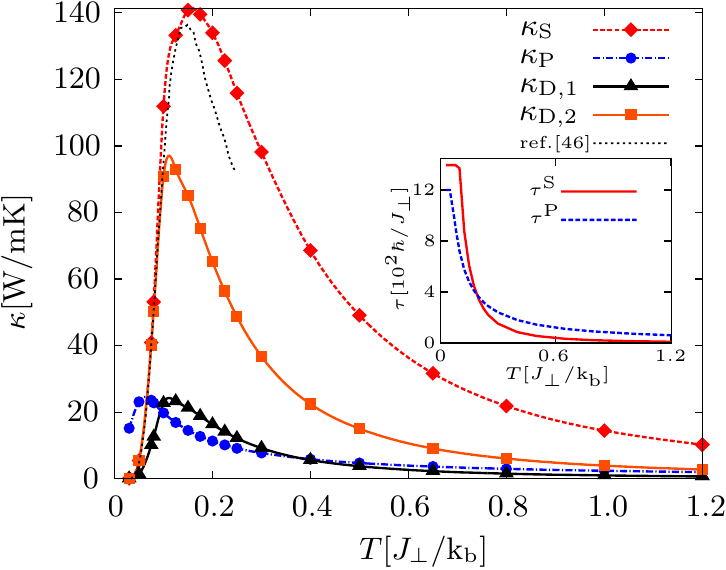}}
\par\end{centering}

\begin{centering}
\subfigure[]{\label{fig3b}
\includegraphics[width=0.8\columnwidth]{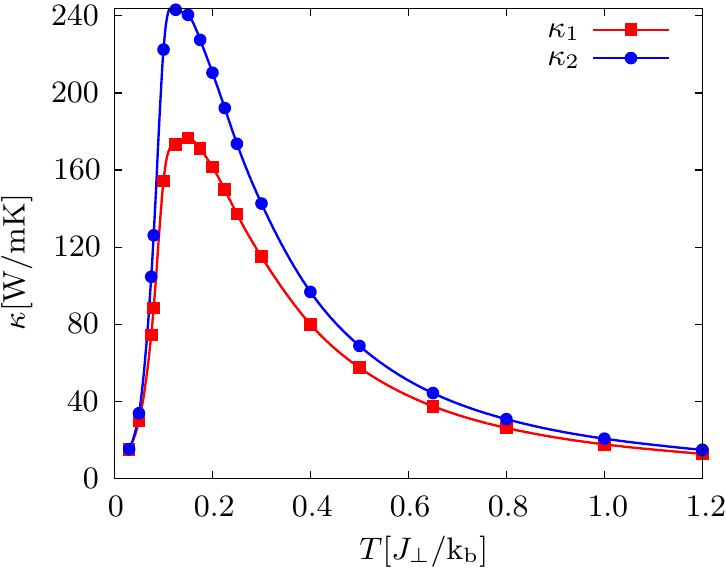}}
\par\end{centering}

\caption{(a) Temperature dependence of thermal conductivities
$\kappa_{\text{S}}$ (checks) and $\kappa_{\text{P}}$ (dots) (\ref{diagcond})
as well as of $\kappa_{\text{D}}$ (\ref{dragcond}) for $g=0.1J_{\parallel}$
(triangles) and $g=0.2J_{\parallel}$ (squares) without magnetic field
$B$. Dashed line without symbols: experimental data for spin heat conductivity in LCCO from ref.\cite{Hess2005a}.
$\kappa_{\text{D}}$ is of the same order of magnitude as $\kappa_{\text{S}}$.
The maximum for $\kappa_{\text{D}}$ lies at lower temperatures than
for $\kappa_{\text{S}}$. 
Inset: Dependence of relaxation times $\tau^{\text{S}}$
(solid) and $\tau^{\text{P}}$ (dotted) in $[10^{2}\hbar/J_{\perp}]$
on $T$ in $[J_{\perp}/\text{k}_{\text{b}}]$ Parameters: $J_{\perp}=\text{k}_{\text{b}} 1200\, \text{K}$,
$J_{\parallel}=0.45J_{\perp}$ (i.e., gap $\approx0.3J_{\perp}$ for
$B=0$), $\omega_{D}\approx0.3\, J_{\perp}$, $a=4\cdot10^{-10}\ \text{m},N=400$.
(b) Total conductivity $\kappa_{\text{S}}+\kappa_{\text{P}}+\kappa_{\text{D}}$
for $g=0.1J_{\parallel}$ ($\kappa_{1}$) and $g=0.2J_{\parallel}$
($\kappa_{2}$).}
\end{figure}

To evaluate the thermal conductivities semi-quantitatively we have
to fix the parameters $g$, $\tau^{\text{S}}$ and $\tau^{\text{P}}$. The diagonal
scattering times $\tau^{\text{S}}$ and $\tau^{\text{P}}$ are chosen
to approximately account for the known transport data of LCCO, \emph{discarding}
any drag contribution a priori. To this end we assume that the low-temperature
scattering for phonons and triplons can be expressed by a temperature-independent
effective mean free path $l_{\text{P(S)}}^{0}$ resulting from grain-boundaries,
impurities etc.. The relation between the mean free paths and the
scattering times is set to $\tau^{\text{P(S)}}(T)=l_{\text{P(S)}}(T)/v_{\text{P(S}\text{),max}}$,
where $v_{\text{P(S)},\text{max}}$ is the maximum phonon(triplon) group
velocity. At high temperatures the diagonal mean free paths are assumed
to be determined by scattering from a bosonic bath, which may either
stem from phonons or triplons. This suggests an expansion of the mean
free path as $l_{\text{P(S)}}^{T}=a_{\text{P(S)}}/T+b_{\text{P(S)}}/T^{2}+\ldots$ \cite{Alvarez2002a}.
Altogether we have $l_{\text{P(S)}}(T)=\mathrm{min}(l_{\text{P(S)}}^{0},l_{\text{P(S)}}^{T})$.
To mimic the experimental data, existing for LCCO for temperatures
$T\lesssim300\,\text{K}$ \cite{Hess2001a,Hess2005a,Hess2006a,Hess2007a,Alvarez2002a},
we found that $l_{\text{P}}^{0}\approx770\text{\AA}$, $a_{\text{P}}\approx55.3\cdot10^{3}\text{\AA} \text{K}$,
and $b_{\text{P}}\approx0$ for the phonons and $l_{\text{S}}^{0}\approx2960\text{\AA}$,
$a_{\text{S}}\approx0$, and $b_{\text{S}}\approx4.18\cdot10^{7}\text{\AA} \text{K}^{2}$ is acceptable.
We emphasize, that, while phonon relaxation mechanisms are rather well
understood, the approximate form of the temperature dependence for
$\tau^{\text{S}}$ merely serves as a fitting parameter, and in particular
$a_{\text{S}}\approx0$ may not apply for $T\gtrsim300\,\text{K}$. However, this
does not impair any of our following conclusions. Finally, to evaluate
the bulk conductivity, we rescale our 1D theory by the perpendicular
surface area $A\approx35.6\text{\AA}^{2}$ per spin ladder.

Using the preceding scattering rates, Fig.~\ref{fig3a} displays
the diagonal conductivities versus temperature at zero magnetic field.
The inset of Fig.~\ref{fig3a} shows $\tau^{\text{P(S)}}$. First,
and as a consequence of our choice of $\tau^{\text{P(S)}}$, the overall
magnitude and peak location of $\kappa_{\text{S(P)}}$ agree roughly
with experiment \cite{Hess2001a,Hess2005a,Hess2006a,Hess2007a,Alvarez2002a}.
Second, Fig.~\ref{fig3a} depicts the drag conductivities, using
the same $\tau^{\text{P(S)}}$, and two typical spin phonon coupling constants
$g$=0.1 and 0.2$J_{\parallel}$. As discussed in section \ref{sub:Spin-Phonon-Coupling},
such values of $g$ are well within reach for cuprates. In turn, as
one main point of this paper, Fig.~\ref{fig3a} demonstrates, that
for reasonable magnetoelastic coupling strengths, the magnitude of
the drag conductivity may well be comparable with that of the diagonal
triplon conductivity. Due to $\kappa_{D}\propto g^{2}$, more quantitative
considerations will depend sensitively on material details, suggesting
that existing interpretations of experimental data in terms of diagonal
conductivities only may need to be reexamined.

Turning to the temperature dependence in Fig.~\ref{fig3a}, and
following the discussion of $G_{\text{S,P,D}}$(T), we first note that $\kappa_{\text{D}}\propto T\tau^{\text{P}}\tau^{\text{S}}$
at high temperatures. Therefore, and as another main point of this
paper, $\kappa_{\text{S}}$ and $\kappa_{\text{D}}$ behave asymptotically identical
at high temperatures, where $\tau^{\text{P}}\propto1/T$ is firmly established.
This can been seen clearly in Fig.~\ref{fig3a}. I.e., the triplon
and drag contributions to the high-temperature transport are entangled
\emph{inseparably}.{\small{} Regarding low temperatures, }$\kappa_{\text{S}}$
displays an Arrhenius-behavior, as expected from the spin gap. This
behavior also translates into the drag. The low temperature phonon
conductivity is $\propto T$. This is an artifact of our 1D phonon
model.

Fig.~\ref{fig3b} shows the total heat conductivity. Three comments
are in order. First, and as a detail of our model parameters, this
figure clarifies, that our choice for the Debye temperature and the
size of the spin gap do \emph{not} fully reflect the material specifics
of LCCO, where the phonon peak is split off into the spin-gap \cite{Hess2001a}.
Second, and more important, this figure demonstrates that even in
the case of strong drag, i.e., $g$=0.2, the drag does not lead to
additional structures in the temperature dependence of the total conductivity.
Finally, since our starting point has been a fit to the observed diagonal
conductivities without considering drag, the magnitude of the total
conductivity does not agree with experiment.

\begin{figure}[tb]
\centering{} \includegraphics[width=0.8\columnwidth]{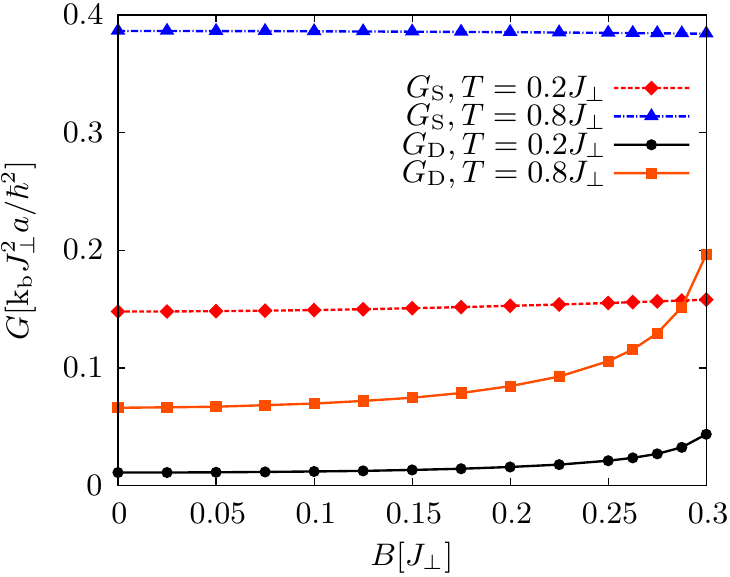} \caption{\label{fig4}Reduced thermal conductivities $G_{\text{S}}$ and $G_{\text{D}}$
from eqn. (\ref{g}) as a function of the magnetic field $B$ for
a temperature smaller ($T=0.2J_{\perp}$) and a temperature larger
($T=0.8J_{\perp}$) than the spin gap. The drag contribution $G_{\text{D}}$
shows a significantly stronger $B$-dependence than $G_{\text{S}}$,
especially for high temperatures. Parameters: $J_{\parallel}=0.45J_{\perp}$
(i.e., gap $\approx0.3J_{\perp}$ for $B=0$), $\omega_{D}\approx0.3\, J_{\perp},N=400$.}
\end{figure}

While the typical exchange energies in LCCO are too large to allow
for sizeable manipulation of the triplon dispersion by experimentally
accessible magnetic fields, it is nevertheless interesting to investigate
also the field dependence of the reduced thermal conductivities $G_{\text{S}}$
and $G_{\text{D}}$. This may be relevant for appropriate low-$J_{\perp,\parallel}$
spin ladders as, e.g., in ref. \cite{Thielmann09a}. The results are
shown in Fig.~\ref{fig4} up to fields of $0.3J_{\perp}$, which almost
close the spin gap for $J_{\parallel}=0.45J_{\perp}$, and for two
temperatures $T=0.2J_{\perp}$ and $0.8J_{\perp}$, which are below
and above the magnon gap. We note that $G_{\text{P}}$ is strictly
field independent in our model. As is obvious from the figure, the
diagonal triplon contribution is almost field independent for the
temperatures depicted. This can be understood in terms of a near cancellation
of the \emph{change} in contributions to $\kappa_{\text{S}}$ from the $S^{z}=\pm1$
triplon branches in finite fields. At significantly lower temperatures
a weak field dependence of $G_{\text{S}}$ can be observed. In contrast
 to $G_{\text{S}}$, $G_{\text{D}}$ shows a clearly visible upward curvature
versus $B$ for larger fields. This behavior increases with temperature.
It is tempting to speculate, that such behavior may help to identify
materials with sizeable drag contributions to their thermal conductivity.

\section{Conclusion\label{sec:Conclusion}}

To summarize, we have investigated the spin-phonon drag effect in
the thermal transport of two-leg spin ladders. Using a bond-operator
description for the spin excitations and magnetoelastic coupling to
longitudinal lattice vibrations we have employed the TCL projection
operator method to map the quantum dynamics of the coupled spin and
phonon system onto a linear Boltzmann equation from which an expression
for the thermal conductivity has been obtained via the Chapman-Enskog
approach. We have shown, that for magnetoelastic coupling strengths
which are well within reach for cuprate spin ladders, the drag contribution
to the total conductivity is \emph{not} negligible, but rather can be
of a size similar to that of the diagonal magnetic heat conductivity.
Moreover, we have found, that the drag may be delicate to discriminate
from the diagonal magnetic conductivity by means of the temperature
dependence, since their high temperature behavior is asymptotically
identical and at intermediate temperatures the drag does not lead
to additional structures. Finally, we have pointed out, that finite
magnetic fields may signal drag contributions to the conductivity.
While the main conclusions of this work should be rather insensitive
to details of the model parameters used in our numerical analysis,
it seems highly desirable to enhance upon the ideas put forward here
by considering a more realistic and 3D version of the phonon spectrum
and the corresponding magnetoelastic coupling geometries.
\begin{acknowledgments}
We thank A. L. Chernyshev, P. Prelov\v{s}ek, and X. Zotos for
helpful discussions. Part of this work has been supported by the DFG
through FOR912 Grant No. BR 1084/6-2, the EU through MC-ITN LOTHERM
Grant No. PITN-GA-2009-238475, and the NTH within the School for Contacts
in Nanosystems. Part of this work has been done at the Platform for
Superconductivity and Magnetism, Dresden (W. B.).\end{acknowledgments}

\end{document}